\documentclass{sig-alternate}

\usepackage{url}
\usepackage{alltt}              
\usepackage[alwaysadjust]{paralist}   

\begin{document}

\global\boilerplate={This work is licensed under the Creative Commons Attribution-Noncommercial-No Derivative Works 3.0 Unported license. You are free to share this work (copy, distribute and transmit) under the following conditions: attribution, non-commercial, and no derivative works. To view a copy of this license, visit http://creativecommons.org/licenses/by-nc-nd/3.0/}

%
\conferenceinfo{DigCCurr2009}{April 1-3, 2009, Chapel Hill, NC, USA}
\global\copyrightetc{\includegraphics[scale=0.7,clip=false]{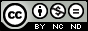}}

\title{Everyone is a Curator:\\
Human-Assisted Preservation for ORE Aggregations}

\numberofauthors{3} 
\author{
\alignauthor
Frank McCown\titlenote{This work performed while Dr. McCown was working with the Digital Library Research \& Prototyping Team at the Los Alamos National Laboratory.}\\
       \affaddr{Computer Science Department}\\
       \affaddr{Harding University}\\
       \affaddr{Searcy, AR 72149}\\
       \email{fmccown@harding.edu}
\alignauthor
Michael L. Nelson\\
       \affaddr{Computer Science Department}\\
       \affaddr{Old Dominion University}\\
       \affaddr{Norfolk, VA 23529 }\\
       \email{mln@cs.odu.edu}
\alignauthor Herbert Van de Sompel\\
       \affaddr{Digital Library Research \& Prototyping Team}\\
       \affaddr{Los Alamos National Lab}\\
       \affaddr{Los Alamos, NM 87545}\\
       \email{herbertv@lanl.gov}
}

\maketitle
\begin{abstract}
The Open Archives Initiative (OAI) has recently created the Object Reuse and Exchange (ORE) project that defines Resource Maps (ReMs) for describing aggregations of web resources.  These aggregations are susceptible to many of the same preservation challenges that face other web resources.  In this paper, we investigate how the aggregations of web resources can be preserved outside of the typical repository environment and instead rely on the thousands of interactive users in the web community and the Web Infrastructure (the collection of web archives, search engines, and personal archiving services) to facilitate preservation. Inspired by Web 2.0 services such as digg, deli.cio.us, and Yahoo! Buzz, we have developed a lightweight system called ReMember that attempts to harness the collective abilities of the web community for preservation purposes instead of solely placing the burden of curatorial responsibilities on a small number of experts.
\end{abstract}




\section{Introduction}

The Web continues to be one of the most useful constructs to disseminate information, enable mass communication, and document our lives. There are, however, two notable challenges, among many, that confront the Web. The first challenge is curatorial. The Web is very difficult to curate because of its shear size and distributed nature, its lack of editorial control and ephemeral qualities.  Web pages that are here today are often gone tomorrow, and links that were once valid now return 404 responses or material that no longer reflects the original link creator's intent.  The transient nature of the Web has been addressed by a number of parties: web archives like the Internet Archive\footnote{\url{http://www.archive.org/}} store historic snapshots of the Web, search engines like Google make temporarily inaccessible pages available from their caches, and personal archiving tools like Spurl\footnote{\url{http://www.spurl.net/}} and WebCite\footnote{\url{http://www.webcitation.org/}} let users archive individual web pages for viewing at a later time.  Although none of these strategies in isolation are completely effective at fending off link rot, the combined efforts of these services, what we call the Web Infrastructure (WI), provides a layer of preservation which adequately protects a massive number of web resources \cite{Nelson07:Using}.

The second challenge facing the Web is organizational in nature. The Web has previously lacked widely accepted standards to group distinct web resources together into a whole.  There are many times when a resource, like an on-line book, academic publication, or news article, is composed of separate web pages or other web-accessible resources.  Although it is usually easy for humans to determine the boundaries of such aggregate resources, it is problematic for an automated agent to do the same \cite{Dmitriev2008:perceive}.

In response to this challenge, the Open Archives Initiative (OAI) has created the Object Reuse and Exchange (ORE) project which provides standards for defining and discovering aggregations of web resources \cite{Lagoze2007:CompoundInfo}.  An aggregation (sometimes called a compound information object \cite{Lagoze2007:CompoundInfo} or compound document \cite{Dmitriev2008:perceive}) may be composed of text, video, images, and any number of web-accessible, URI-identified resources.  For example, a scholarly publication may consist of an HTML ``splash page'' along with versions of the paper in PDF and PostScript format, a video slideshow, and the raw data used to perform the related research.  An aggregation documenting a special event like 9-11 could be composed of images, video footage, news stories, and blog posts.  Aggregated resources may reside on the same website (e.g., in the same repository), or they may be distributed across the Web.

The ORE Data Model introduces the concept of a Resource Map (ReM), a web resource that describes an aggregation. ReMs act as an organizational unit, defining the boundaries of an aggregation and indicating the relationships between the aggregated resources.  Like their aggregated resources, ReMs have their own URIs.  They may be housed in an institutional or academic repository like arXiv\footnote{\url{http://arxiv.org/}} where they may receive a high degree of monitoring by administrators.  Others may exist outside the repository where they may be maintained by any number of individuals.

Unfortunately, whether ReMs are maintained inside the walls of a repository or outside in the wild, they may eventually fall prey to neglect.  ReMs share many of the same preservation difficulties as other web resources: ReMs may change over time, move to different URLs, or disappear completely from the Web.  However, they also present additional challenges because the resources they aggregate may also change, move to different URLs, or disappear.  This added dimension suggests ReMs may require more curatorial attention than other web resources.

In this paper, we explore some strategies that can free the ReM creator from the burden of full curatorial responsibilities and instead distribute or democratize the workload to the masses.  Inspired by Web 2.0 sites that rely on the public for producing and maintaining content, we have developed a system called ReMember which leverages the distributed efforts of the public who interact with web archives, search engines, and personal archiving services (the Web Infrastructure) to maintain the integrity and accuracy of ReMs.  Employing a small number of experts to provide equivalent curatorial services would be prohibitively expensive and would not scale to the Web.  But by distributing the effort to the public, we believe the small, contributed efforts of many in conjunction with the WI will allow us to curate ReMs on the scale of the Web.

\section{Background}

\subsection{Object Resource and Exchange}

As mentioned earlier, humans can easily determine the boundaries of an aggregation, but it is very difficult for a machine to do the same.  Sharing the Semantic Web's goal of enabling a machine-readable Web and the Linked Data vision of connecting disparate datasets, the ORE project aims to create standards that allow aggregations of web resources to be defined and discovered \cite{Lagoze2008:OREPrimer,Lagoze2008:ObjectReUse,Sompel07:Interoperability}.  These aggregations are conceptual resources which are made concrete by Resource Maps (ReMs).  ReMs enumerate the aggregated resources (ARs) that make up an aggregation and include descriptive metadata about each AR.  Aggregations and ReMs have distinct URIs, but dereferencing an aggregation's URI will lead to the authoritative (or trusted) ReM that describes it.

\begin{figure*}
\begin{center}
\includegraphics[scale=0.75,clip=false]{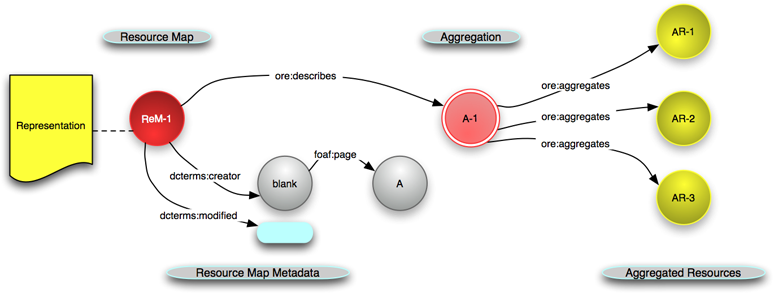}
\caption{An example Resource Map, borrowed from [7].}   
\label{fig:example-rem}
\end{center}
\end{figure*}

An example ReM is shown in Figure \ref{fig:example-rem} where ReM-1 is the URI identifying the ReM, and A-1 is the URI identifying the aggregation.  The aggregation contains three aggregated resources (AR-1, AR-2, and AR-3), and RDF triples are used to describe the relationships between the ReM, aggregation, and ARs.

ReMs may be serialized in a number of formats like RDF/ XML and RDFa, but the simplest format is the Atom Syndication Format \cite{rfc4287}.  Atom is a popular syndication format for blogs, but increasingly it has been used for other purposes like the Google Data API\footnote{\url{http://code.google.com/apis/gdata/overview.html}}.

\begin{figure}
\begin{small}
\begin{alltt}<?xml version="1.0" encoding="utf-8"?>
<feed xmlns="http://www.w3.org/2005/Atom">
  <id>http://arxiv.org/rem/astro-ph/0601007#aggregation</id>
  <link
    href="http://arxiv.org/rem/astro-ph/0601007v2"
    rel="self" type="application/atom+xml"/>
  <category scheme="http://www.openarchives.org/ore/terms/"
    term="http://www.openarchives.org/ore/terms/Aggreagation"
    label="Aggreagation"/>
  <link href="http://arxiv.org/rem/astro-ph/0601007"
    rel="self" type="application/atom+xml"/>
  <title>Parametrization of K-essence
    and Its Kinetic Term</title>
  <author><name>Hui Li</name></author>
  <author><name>Zong-Kuan Guo</name></author>
  <author><name>Yuan-Zhong Zhang</name></author>
  <updated>2007-10-10T18:30:02Z</updated>
  <entry>
    <id>tag:arxiv.org,2007:astro-ph/0601007v2:ps</id>
    <link
      href="http://arxiv.org/abs/astro-ph/0601007"
      rel="alternate" type="text/html"/>
    <title>Splash Page for "Parametrization of
      K-essence and Its Kinetic Term"</title>
    <updated>2006-05-31T12:52:00Z</updated>
  </entry>
  <entry>
    <id>tag:arxiv.org,2007:astro-ph/0601007v2:pdf</id>
    <link
      href="http://arxiv.org/pdf/astro-ph/0601007v2"
      rel="alternate" type="application/pdf"/>
    <title>PDF Version of "Parametrization of
      K-essence and Its Kinetic Term"</title>
    <updated>2006-05-31T12:52:00Z</updated>
  </entry>
...
</feed>\end{alltt}
\end{small}
\caption{A Simple Resource Map for an arXiv e-print.}
\label{fig:example-arxiv-rem}
\end{figure}


An example ReM\footnote{The example uses version 0.9 of ORE which has recently been replaced by version 1.0. Although the Atom serialization in version 1.0 is significantly different, it can easily be implemented in ReMember without any impact on its functionality.} for an arXiv e-print is shown in Figure \ref{fig:example-arxiv-rem}.  This simple example shows two ARs, an HTML ``splash page'' and a PDF, that, together with other ARs not shown in the example, constitute the e-print resource.  The ReM itself was last updated on 2007-10-10, but an updated timestamp of 2006-05-31 was used for the ARs; this later timestamp may reflect when the ARs were last modified or when the Atom entry was last modified.  Both ARs have link elements which indicate the URLs of the resources.

ReMs may be created by anyone.  They may be discovered by humans and bots by following a link to them.  An HTML resource that is aggregated by a ReM may also contain a link element which points to the ReM; non-HTML ARs can make use of the HTTP link response header which performs the same function.  Batch discovery methods like SiteMaps and OAI-PMH may also be used to discover ReMs.  More technical details of ReM Atom serialization and other ORE standards are available on the OAI-ORE website\footnote{\url{http://www.openarchives.org/ore/toc}}.

\subsection{Web Infrastructure}

The Web Infrastructure (WI) is the collective activities of web archives (e.g., Internet Archive), search engines (e.g., Google, Live Search, and Yahoo), personal archiving tools (e.g., Spurl, Hanzo:web, and WebCite), and research projects (e.g., CiteSeerX  and NSDL) that refresh and migrate large amounts of web content as by-products of their primary services \cite{Nelson07:Using}.  The WI can be used as a passive service for a number of preservation functions.  For example, websites that have been lost without backups can be reconstructed from the WI using Warrick \cite{McCown2006:Lazy}, and web resources that move from one URI to another can be relocated using Opal \cite{Harrison:Just-In-Time}.

\begin{figure}
\begin{center}
\includegraphics[scale=0.3,clip=false]{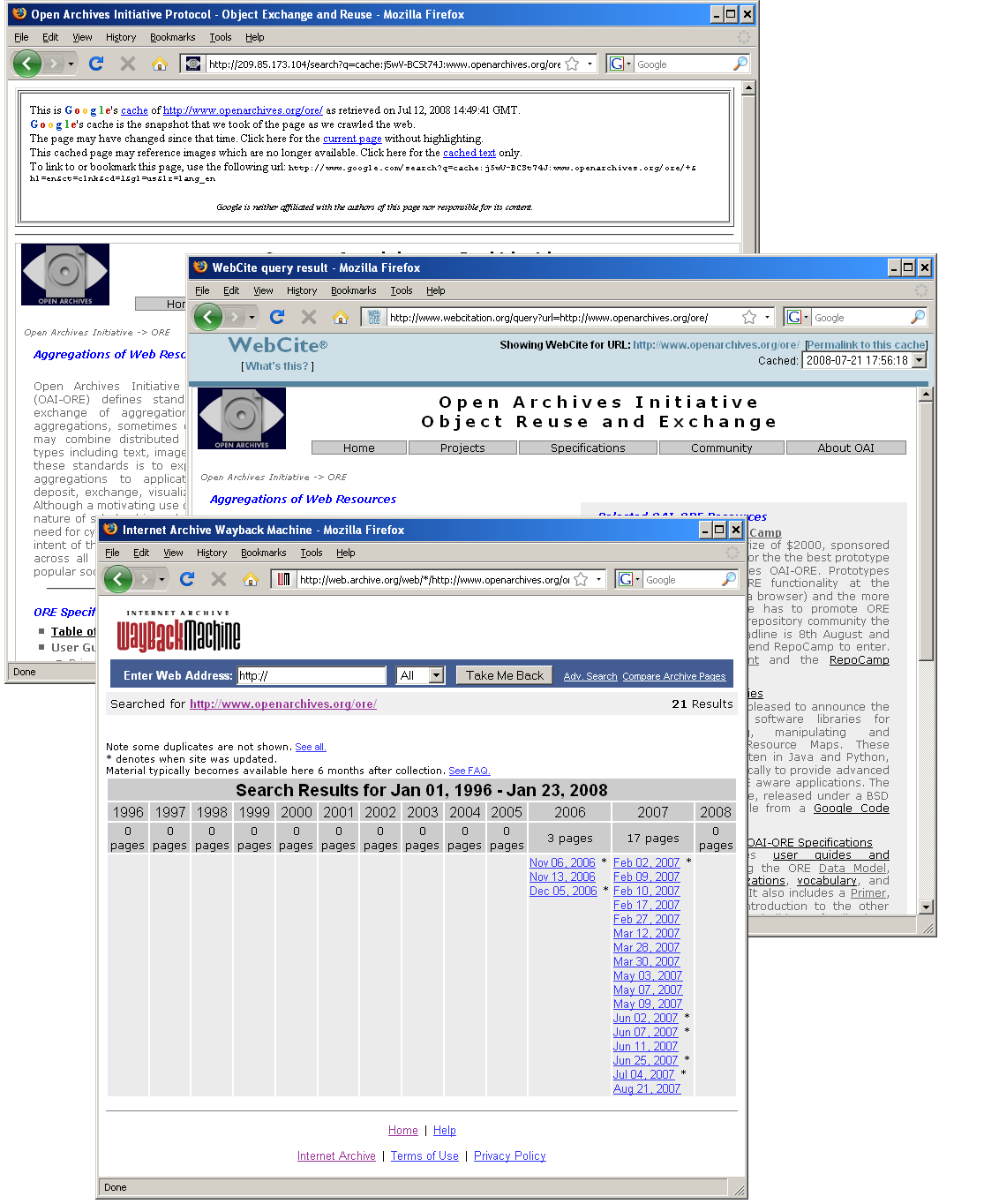}
\caption{Copies of the ORE web page at http://www.openarchives.org/ore/ stored in Google's cache, the WebCite archive, and the Internet Archive.}
\label{fig:wi-screenshot}
\end{center}
\end{figure}

Figure \ref{fig:wi-screenshot} illustrates how the WI has captured multiple versions of the ORE home page.  The upper-left screenshot shows Google's cached copy of the page as they crawled it on 2008-07-12.  The middle screenshot shows WebCite's archived version of the web page from 2008-07-21 (as initiated by one of the authors), and the bottom screenshot shows multiple versions of the same page available from the Internet Archive (IA) from 2006-11-06 to 2007-08-21.  Unfortunately, IA has a 6-12 month lag in their archive, so they do not have any copies available from 2008.

There are some notable differences between various members of the WI.  Search engines usually have the most up-to-date and widest breadth of resources available from their caches because of the competitive nature of web search and huge investments in web crawling infrastructure \cite{McCown07:Characterization}. However, when search engines discover that a web resource has been changed, they discard the old version of the resource for the new one.  They also migrate textual resources like PDF and Microsoft Word into HTML pages that lose their formatting and embedded images.

Web archives like IA also rely primarily on web crawling to discover web resources.  IA keeps old versions of resources in their original format, but as stated before, they are slow to update their archive, and their snapshot of the Web may not be as broad or complete as the search engines.  Personal archiving services like WebCite \cite{Eysenbach05:Going} require a user to initiate the archival process rather than relying on web crawling for content.  Therefore resources that may not have been deemed important will likely be missed by such a service.  IA and WebCite both store content in its original format, a significant advantage over search engines.

\section{Client-Assisted Preservation}

\subsection{Overview}

\begin{figure}
\begin{center}
\includegraphics[scale=0.68,clip=true,viewport=0 0 370 312]{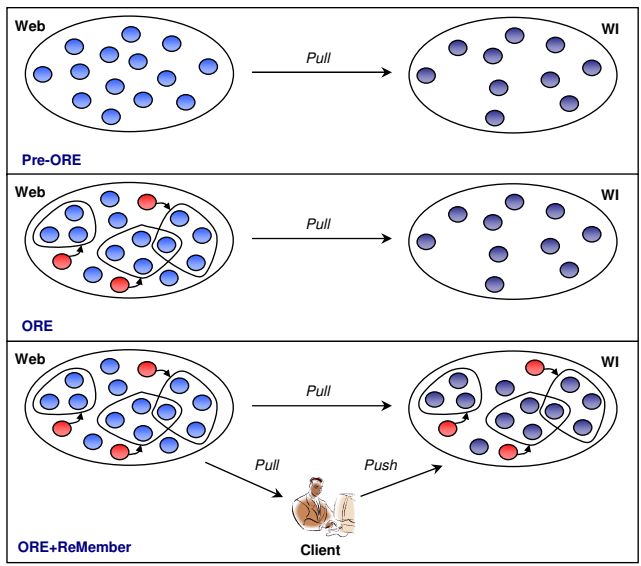}
\caption{Pre-ORE (top) shows the WI migrating and refreshing un-aggregated resources from the Web (for simplicity, links between resources are not shown). With the introduction of ORE (middle), aggregated resources are delineated by Resource Maps (red). With the introduction of ReMember (bottom), humans (clients) are leveraged to push aggregated resources and ReMs into the WI.}
\label{fig:ore-remember}
\end{center}
\end{figure}

Several of the WI members like the Internet Archive and commercial search engines rely primarily on web crawling to populate their repositories.  This automated ``pull'' methodology, illustrated on the top pane of Figure \ref{fig:ore-remember}, works well in terms of finding a large number of web resources, but some may be missed (e.g., resources that require too many hops from the root page,  resources for which links may not be found, and ``unpopular'' resources).

With the introduction of ORE (middle of Figure \ref{fig:ore-remember}), ReMs (red dots) delineate the aggregated resources on the Web (for simplicity, aggregations and aggregations' ReMs are shown as a single unit).  The WI may archive and cache some ReMs and ARs as they do any web-accessible resource.  In previous work, we showed how IA could archive evolving ReMs and their evolving ARs without any architectural changes \cite{VandeSompel2007:CompoundDemo}.

By introducing our ReMember system, we hope to enable millions of Web users to perform a small amount of curatorial work in keeping ReMs that change over time current by ensuring that all AR URIs resolve at various points in time to the correct content.  Individuals will push ReMs and ARs into the WI, allowing for more complete and accurate coverage than what is now possible (bottom of Figure \ref{fig:ore-remember}).

\subsection{Architecture}

In order to harness the curatorial power of the masses, we have constructed the ReMember prototype which adheres to several important design goals:
\begin{enumerate}
  \item Resource producers should easily enable inclusion of their resources and ReMs into the system.
  \item The system should rely on the WI for storage since we do not personally have sufficient storage capacity for potentially millions of resources.
  \item The system should not assume any WI member will always be accessible.  Members of the WI may come and go, and therefore copies of the resources should be spread throughout the WI.
  \item The system should help users relocate missing resources by maintaining a small fingerprint that might help identify the resource.
  \item Changes to ReMs should be logged over time to allow for rollback operations.

\end{enumerate}

ReMember is a lightweight system that attempts to put as little demand as possible on resource producers and consumers.  In order for resource producers or maintainers to mark their ARs and ReMs for inclusion in ReMember, they may insert an HTML snippet into the bottom of their HTML page that produces a ``Preserve this Object'' link.  When a user who is viewing the page clicks on the link, the user will be prompted to preserve the ReM and its ARs.  If AR maintainers are unable or unwilling to add an HTML snippet to their resources, users may still curate ReMs and their ARs by use of a browser plug-in (to be implemented) that automatically discovers ReMs and facilitates submitting them to ReMember.  Individuals may also submit ReMs to ReMember  directly using its web interface.

In the future, we envision a del.icio.us-like interface that displays the ReMs that users are most frequently preserving.  But unlike del.icio.us, we may be more interested in showing users the resources that are \emph{not} being preserved, the unpopular ReMs, since these resources are in most need of the community's attention.

\begin{figure}
\begin{center}
\includegraphics[scale=0.77,clip=true,viewport=0 0 310 180]{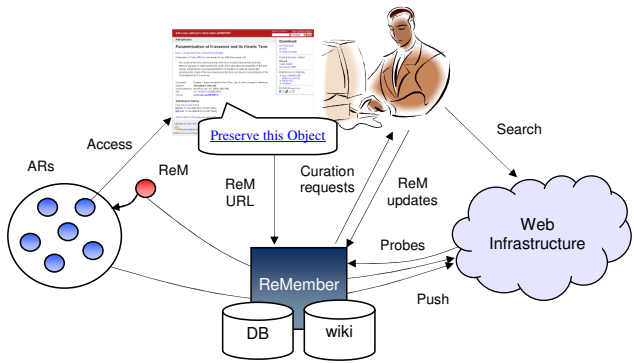}
\caption{Diagram showing the interaction between resource producers, resource consumers, and ReMember.}
\label{fig:remember-diagram}
\end{center}
\end{figure}

The architectural and process overview of ReMember is shown in Figure \ref{fig:remember-diagram}.  When a user who accesses an aggregated resource clicks on the ``Preserve this Object'' link, the AR's ReM URL is submitted to ReMember.  If the ReM has never been seen by ReMember, it will immediately push a copy of the ReM and each AR (obtained from the ReM) to the WI via a personal archiving service (WebCite).  ReMember will also create a lexical signature (five words that could be used to uniquely identify the resource when searching the Web \cite{phelps:robust}), store the ReM in its wiki (for version control), and store a thumbnail image snapshot of each AR.  The lexical signature and thumbnail, together with the URL and ReM AR metadata (title, author, description, etc.), will act as a fingerprint for an AR in case the WI were to lose its copies or if the user needed to find an AR that moved to a different URL.  If the user is willing, he/she will also submit the URL to several search engines if they have not yet indexed the AR (this usually requires the user to solve a CAPTCHA and thus cannot be fully automated).

Subsequent accessing of the ReM in ReMember will prompt the user to correct any broken links to missing ARs.  This involves showing the user any older copies of the AR that may be found in the WI using the AR's old URL.  The user can also search the Web for the new location of the missing AR using the metadata from the ReM and the lexical signature.  The user will also be prompted to examine any ARs that have changed since the last time they were archived in the WI; significant changes warrant re-archiving, obtaining new lexical signatures, and creating a new thumbnail.  Any changes made to the ReM are archived to the WI, and the changes are noted in the wiki.  By ensuring that the ReM is valid each time a user visits ReMember, we may allow users to view older versions of the ReM and its associated ARs at various points in time by pointing to archived versions of ARs in the WI.

The following summarizes the data being stored in ReMember and the WI:\\

\textbf{Stored in ReMember}
\begin{enumerate}
\begin{compactenum}
  \item ReM at time $t_{i}$ (document)
  \item For each AR$_{j}$ in ReM at $t_{i}$ (URI)
\begin{itemize}
\begin{compactenum}[1.]
  \item Metadata (title, author, etc.)
  \item Lexical signature
  \item Image thumbnail
  \item URI of AR$_{j}$ in WI
\end{compactenum}
\end{itemize}
\end{compactenum}
\end{enumerate}

\textbf{Stored in the WI}
\begin{enumerate}
\begin{compactenum}
  \item ReM at $t_{i}$ (document)
  \item $AR_{j}$ at $t_{i}$ (document)
\end{compactenum}
\end{enumerate}

\vspace{1 mm}  

\subsection{Possible Scenarios}

\begin{figure}
\begin{center}
\includegraphics[scale=0.9,clip=true,viewport=0 0 250 290]{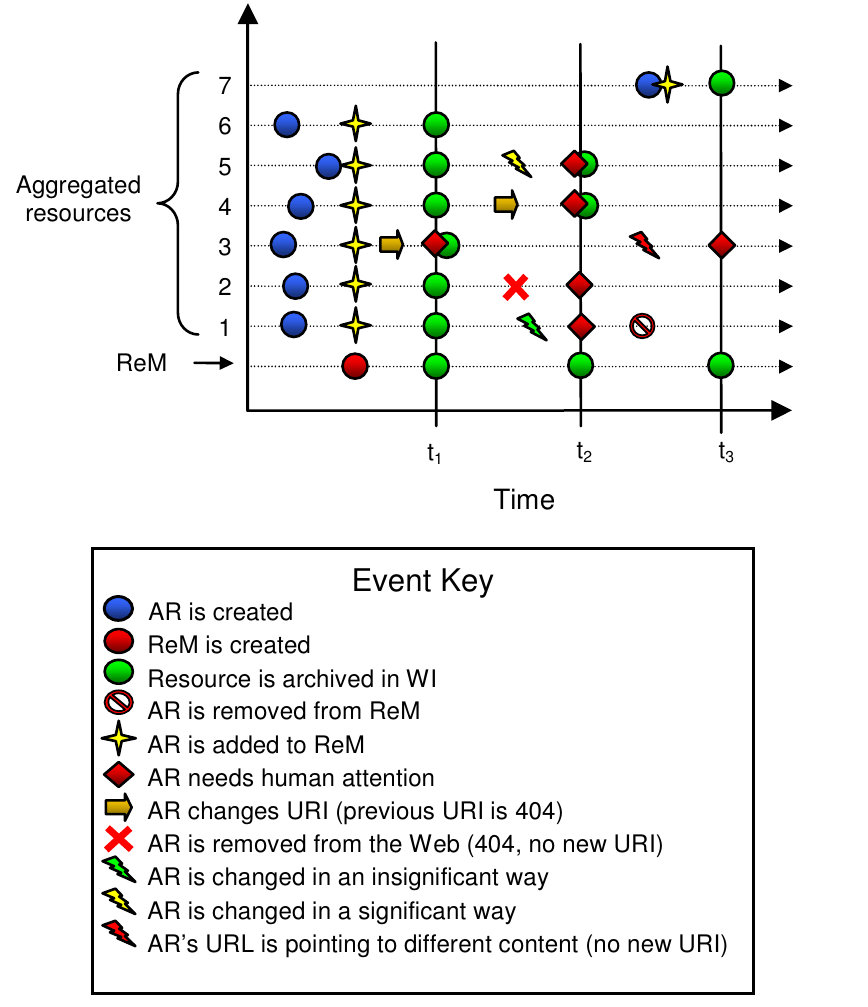}
\caption{Diagram showing various events in time which impact Aggregated Resources and Resource Maps.}
\label{fig:scenarios}
\end{center}
\end{figure}

A ReM and its ARs might exhibit a variety of changes over their lifetimes as illustrated in Figure \ref{fig:scenarios}.  The vertical lines at $t_1$, $t_2$, and $t_3$ represent users accessing ReMember at various times and curating the ReM.  The diagram is not exhaustive, but the most common events are accounted for.  Figure 6 shows six ARs (1-6) that are created before the ReM.  The ARs in Figure \ref{fig:scenarios} are added to the ReM by the ReM creator at creation time.  AR7, which is not created until some time later, is also added to the ReM before time $t_3$ by the ReM's maintainer.

Figure \ref{fig:scenarios} shows AR3 moving from one URI to another before anyone has had a chance to access the resources through ReMember.  So when one of the resources is accessed at time $t_1$, all the ARs can be archived except AR3; a user is required to investigate where the resource has moved.  At this point, there are two components that ReMember can use to assist the user: the AR's previous URI and the AR's metadata.  ReMember will aid the user in finding the resource by first examining the WI for copies of the resource using the old URI (a search engine's cache or IA).  Having access to old copies may help the user locate the new URI of the AR with the aid of a WI search engine like Google.  Even if old copies are not found, the metadata can be used as a query in a web search.  Once the user finds the resource's new URL, the ReM is updated with the new information, the ReM is archived in the WI, and the changes are logged on ReMember's wiki.

Figure \ref{fig:scenarios} shows that AR4 also changes URIs, but this takes place after ReMember has had the chance to create a lexical signature and thumbnail snapshot of the resource.  This information, along with the metadata stored in the ReM, can be used to help the user at time $t_2$ to relocate the resource.  AR2 has been completely removed from the Web between $t_1$ and $t_2$, so the user will not be able to find a new copy of it at $t_2$.  This requires the user to flag the resource as being no longer accessible on the Web.  The same decision will need to be made for AR3 at $t_3$ when the URI still resolves but points to the wrong content, and the AR is not available at any URI except at WebCite.  AR1 was removed from the ReM outside of the ReMember system sometime between $t_2$ and $t_3$, but user intervention is not required; ReMember need only archive the new ReM and capture the changes in its wiki.

AR1 and AR5 undergo some degree of change between $t_1$ and $t_2$, both of which require a user to decide if the change is significant enough to warrant re-archiving the AR or finding a suitable replacement AR.  AR1 experiences only a minor change, like a change in the date or an advertisement or maybe a minor layout adjustment.  AR5 undergoes a significant change, like an update to a blog entry or new version of an academic paper.  Although heuristics can be devised to determine the degree of change, we believe a human (assisted by the heuristics) is more likely to make the best curatorial decisions.

\section{Example: Academic Bibliography}

To illustrate how ReMember might be used in a real world scenario, we created a ReM based on a bibliography found online\footnote{ \url{http://www.chin.gc.ca/English/Digital_Content/Digital_Preservation/bibliography.html }} that points to twenty-six online papers about digital preservation (the papers were housed on multiple websites).  The web page serves as a human-readable aggregation, but a machine would have difficulty determining which links were to be included in the aggregation and which simply pointed to related websites.  We added the splash page and each of the bibliographic entries to the ReM as aggregated resources.  The title of the papers and authors were entered as metadata for each AR.

When accessing the newly created ReM for the first time in ReMember, the ReM is pushed to the WI (WebCite) and to the wiki.  Each AR is downloaded, and a screenshot thumbnail and lexical signature is created for those ARs that are successfully downloaded.  The ARs are also pushed to WebCite.

\begin{figure}
\begin{center}
\includegraphics[scale=0.4,clip=false]{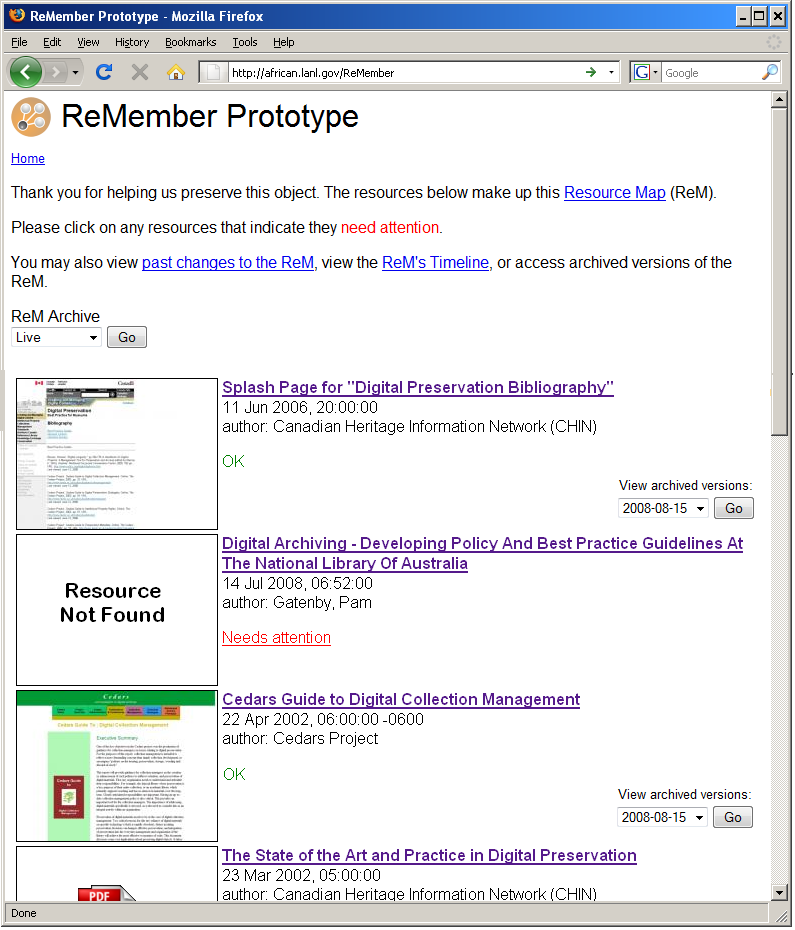}
\caption{Screenshot of ReMember.}
\label{fig:remember-prototype}
\end{center}
\end{figure}

A screenshot of ReMember is shown in Figure \ref{fig:remember-prototype} as the user would see it when accessing the ReM.  Each AR has a thumbnail image shown on the left with its accompanying title, updated timestamp, author, and any other descriptive metadata.

As indicated in the screenshot, the splash page (the first AR) does not need any user attention since this is the first time ReMember has seen this AR, and it was successfully accessed.  The second AR, however, returned a 404 response when ReMember attempted to download it, so the user's assistance is required.  When the user clicks on ``Needs attention,'' a new browser window will appear which will first show any copies of the AR that the WI may have.  Since the resource has been missing from the Web for some time, the search engine caches no longer have a copy, but IA has a version from 2007-01-06.  The user could update the ReM to use IA's version of the resource, but we encourage the user to locate the resource at some other URL instead to keep the ReM pointing to the ``live'' Web as much as possible. We assume the live Web will usually contain the most recent version of the resource.

\begin{figure}
\begin{center}
\includegraphics[scale=0.37,clip=false]{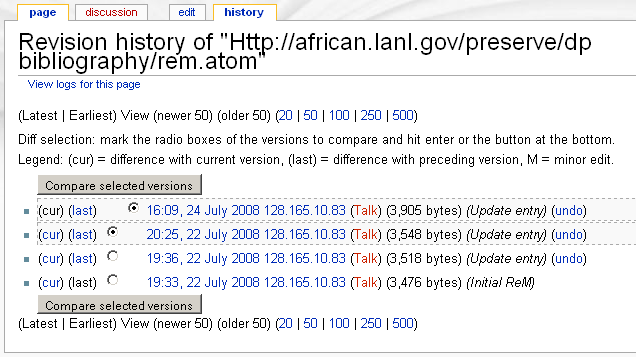}
\caption{Screenshot of ReMember's wiki for ReM version control.}
\label{fig:wiki-screenshot}
\end{center}
\end{figure}

When assisting the user in finding a live version of the resource, ReMember uses the title of the missing article to pre-build the query in a Google search form (Yahoo and Live Search can also be searched). The user may also use other metadata such as author to perform his/her search.  After doing some searching, the user may give up if they were unable to find a new URL for the article, so the user will update the ReM to indicate the resource cannot be found on the live Web.  The ReM's changes are logged in the wiki. A screenshot of the wiki is shown in Figure \ref{fig:wiki-screenshot} after making several changes to the ReM.  Users may view this wiki at any time and can rollback any erroneous modifications that may have been made.

The next time the same ReM is accessed in ReMember, the live ReM will be checked to see if it has undergone any changes since the last time it was archived.  Each AR is also accessed and compared with its archived version in WebCite. If one of the ARs in Figure \ref{fig:remember-prototype} appears to have undergone a change of some sort, ReMember will request the user's attention for this AR, and the user may decide that the newer version should be archived.  ReMember will push a copy of the updated article to WebCite, compute a new lexical signature, and take a new snapshot.

\begin{figure}
\begin{center}
\includegraphics[scale=0.4,clip=false]{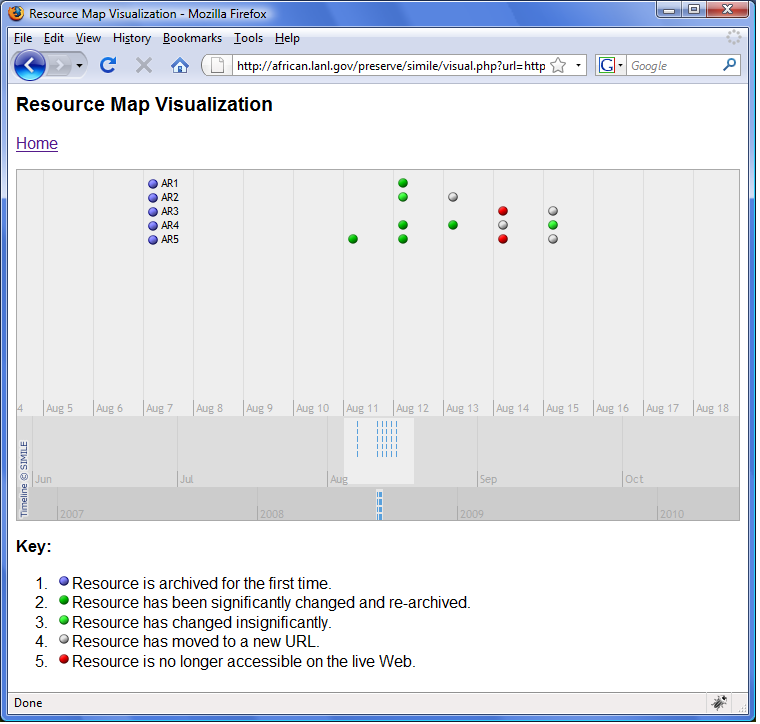}
\caption{Screenshot of ReMember visualization timeline.}
\label{fig:rem-visualization}
\end{center}
\end{figure}

ReMember also allows users to view a timeline of changes that occur to ARs using MIT's Simile Timeline\footnote{\url{http://code.google.com/p/simile-widgets/wiki/Timeline}} widget. Figure \ref{fig:rem-visualization} shows a timeline for the ARs in this case study's ReM. All the resources were archived for the first time on Aug 7, and several ReMs experienced significant and insignificant changes on different days; some moved to new URLs or went missing on the live Web. This visualization helps users see which ARs are the most volatile over time. While ReMember does not attempt to show explicit differences between each version of an AR, other projects like the Past Web Browser do \cite{Jatowt2008:History}.

Appendix A shows several other example ReMs being curated in ReMember.

\section{Ongoing Work and Conclusions}

Link rot has been a continual adversary of the Web, and a number of solutions have been offered to combat the problem (e.g., \cite{beynon2007ghl, Harrison:Just-In-Time, Morishima08:PageChaser}).  Our system does not replace such systems but augments them by attempting to harness the abilities of the web community to curate ReMs when automated processes are not enough.

As ReMs become more prevalent on the Web, we hope ReMember will be embraced by a community of individuals who desire to keep ReMs on particular topics accurate, just as Wikipedia has been embraced by a large community to curate a large number of articles on a variety of topics.  Like Wikipedia, our system will likely be targeted by spammers, and it remains to be seen what editorial controls or techniques will be required to fight mischievous alterations.

We are also investigating how the community could take ownership of a ReM and add and delete ARs from it.  We believe that the community would take more interest in curating ReMs if they could personally enhance its usefulness like one might enhance a Wikipedia article.  Again, spam issues will likely be a significant challenge.

\section{Acknowledgments}

This research is supported in part by the Andrew Mellon Foundation.  We thank members of the LANL Digital Library Research \& Prototyping Team, especially Ryan Chute, for their technical assistance in developing the ReMember prototype.

%

\appendix
\label{appendix-a}
\section{Additional Examples}
Some additional examples are included here to show different types of ReMs being curated with ReMember.

\begin{figure}[!b]
\begin{center}
\includegraphics[scale=0.8,clip=false]{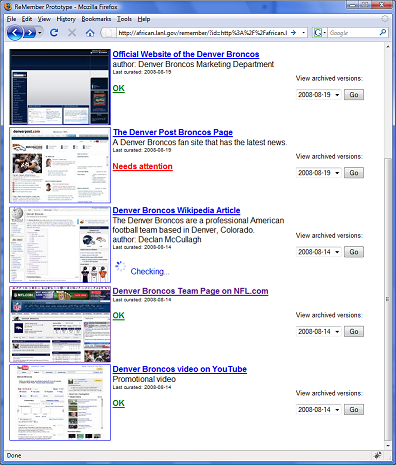}
\caption{Curating a Denver Broncos ReM.}
\label{fig:remember-broncos}
\end{center}
\end{figure}

Figure \ref{fig:remember-broncos} shows a ReM that points to a number of web pages about the Denver Broncos football team. The second AR needs attention because its text has changed significantly since the last time the AR was curated. The third AR is still being checked; ReMember checks resources asynchronously since some web servers may respond slowly to requests.

\begin{figure}
\begin{center}
\includegraphics[scale=0.8,clip=false]{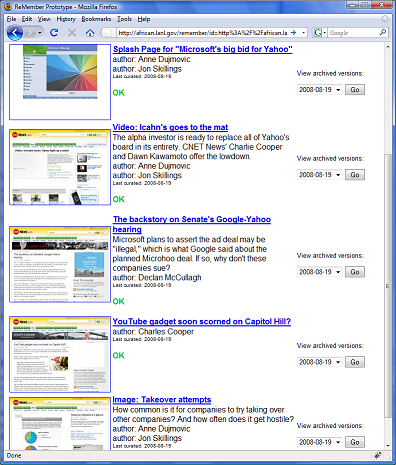}
\caption{Curating a ReM about the proposed 2008 Microsoft-Yahoo merger.}
\label{fig:remember-microhoo}
\end{center}
\end{figure}

\begin{figure}
\begin{center}
\includegraphics[scale=0.3,clip=false]{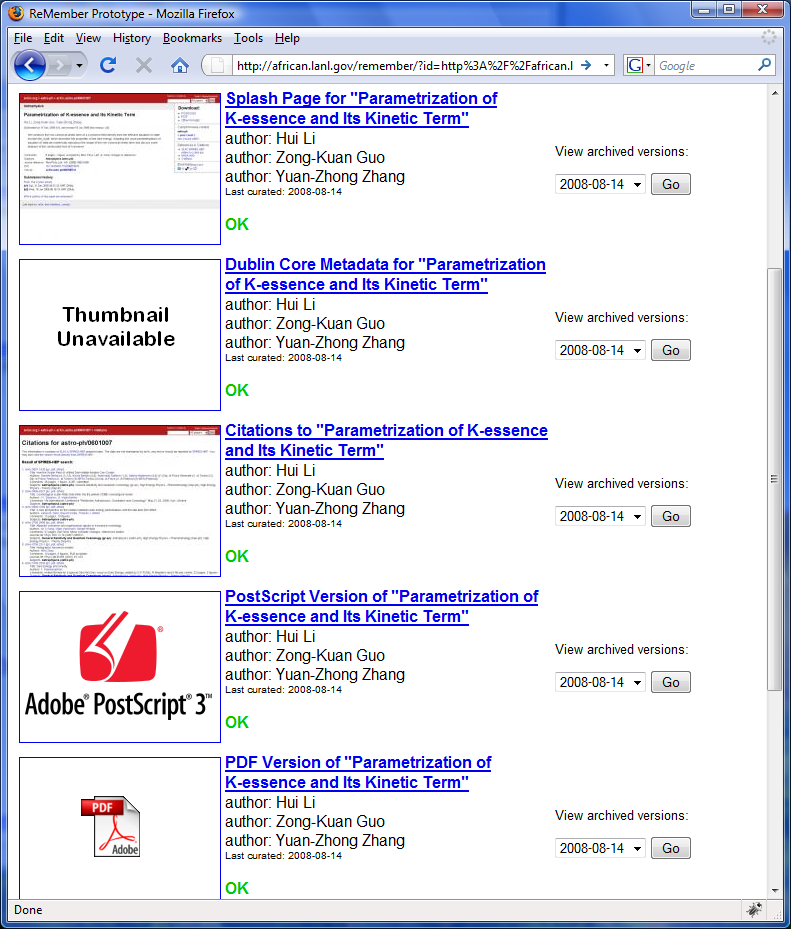}
\caption{Curating an arXiv e-print ReM.}
\label{fig:remember-arxiv}
\end{center}
\end{figure}

A ReM that points to various news reports, images, and videos (all from CNET.com) of the proposed Microsoft-Yahoo merger is shown in Figure \ref{fig:remember-microhoo}. Finally, Figure \ref{fig:remember-arxiv} shows the ReM from the arXiv e-print example of Figure \ref{fig:example-arxiv-rem}.

\end{document}